\documentclass[12pt]{article}

\newcommand{\be}{\begin{equation}}
\newcommand{\ee}{\end{equation}}

\usepackage{amsmath,amsfonts,amsthm,amssymb,eucal}
\usepackage{breqn}

\begin{document}
%%%%%%%%%%%%%%%%%%%%%%%%%%%%%%%%%%%%%%%%%%%%%%%%%%%%%%%%%%%%%

\begin{center}

Mathematical Methods in the Applied Sciences. 2021. Vol.44. No.14. P.11514-11525. 

\vskip 7mm

{\bf \large Predator-Prey Models with Memory and Kicks: } 
\vskip 3mm

{\bf \large Exact Solution and Discrete Maps with Memory} \\

\vskip 7mm
{\bf \large Vasily E. Tarasov}$^{1,2}$ \\
\vskip 3mm

${}^1$ {\it Skobeltsyn Institute of Nuclear Physics, \\ 
Lomonosov Moscow State University, Moscow 119991, Russia} \\
{E-mail: tarasov@theory.sinp.msu.ru} \\

${}^2$ {\it Faculty "Information Technologies and Applied Mathematics", \\ Moscow Aviation Institute (National Research University), Moscow 125993, Russia } \\

\begin{abstract}
In this paper, we proposed new predator-prey models that take into account memory and kicks. 
Memory is understood as the dependence of current behavior on the history of past behavior. 
The equations of these proposed models are generalizations of the Lotka-Volterra and Kolmogorov equations by using the Caputo fractional derivative of non-integer order and periodic kicks. 
This fractional derivative allows us to take into account memory with power-law fading.
The periodic kicks, which are described by Dirac delta-functions, take into account short duration of interaction between predators and prey.
For the proposed equations, which are fractional differential equations with kicks, we obtain exact solutions that describe behaviors of predator and prey with power-law fading memory. 
Using these exact solutions, we derive, without using any approximations, new discrete maps with memory that represent the proposed predator-prey models with memory. 
\end{abstract}

\end{center}

\noindent
MSC: 26A33; 34A08 \\
%%% 26A33 	Fractional derivatives and integrals
%%% 34A08 	Fractional differential equations
PACS: 45.10.Hj; 05.45.-a; 05.90.+m \\
%%% 45.10.Hj Perturbation and fractional calculus methods
%%% 05.45.-a Nonlinear dynamics and chaos 
%%% 05.90.+m Other topics in statistical physics, thermodynamics, and nonlinear dynamical systems

\noindent
Keywords: fractional dynamics; discrete map with memory; 
fractional differential equation; fractional calculus; 
processes with memory; predator-prey model; Lotka-Volterra equations; Kolmogorov equations;

%%%%%%%%%%%%%%%%%%%%%%%%%%%%%%%%%%%%%%%%%%%%%%%%%%%%%%%%%%%%%

\section{Introduction}

Predator-prey models take into account interactions of two or more types of populations. 
These models describe dynamics of populations that interact, thereby affecting each other's growth rates. 
The first predator-prey model was proposed by Alfred J. Lotka \cite{Lotka1910} to describe autocatalytic chemical reactions. 
Then this model was extended to biological systems
by Alfred J. Lotka in \cite{Lotka1925}, 
Vito Volterra in \cite{Volterra}, and
Andrey N. Kolmogorov in \cite{Kolmogorov1936} (see also \cite{Kolmogorov1972,Sigmund}). 

The predator-prey models are actively used not only in biology, but also in economic sciences. For example, the predator-prey economic model was proposed by Paul A. Samuelson \cite{Samuelson}, who is the first American to win the Nobel Memorial Prize in Economics. 

Mathematically, the Lotka-Volterra and Kolmogorov models are represented by a system of two nonlinear first-order differential equations. These equations describe the dynamics of biological (or economic) systems, in which two species interact: one as a predator and the other as a prey. 
The Lotka-Volterra equations can be considered as a special case of equations of the Kolmogorov model, which describes dynamics of systems with predator-prey interactions, competition, disease, and mutualism.

It is known that the differential equations of first-order and
any integer-order cannot take into account memory effects.
Memory is understood as a property of processes that describes the dependence of the current state of the process on the history of state changes in the past time interval (for example, see \cite{Concept}). 

Adaptations developed by prey to counter predators facilitate the development of mechanisms by predators to overcome these adaptations. 
In general, these adaptation processes can be based on the presence of some type of memory in prey and predators. 

The description of life, social and economic processes should take into account that the behavior of actors may depend on the history of previous changes in these processes.
It is strange to neglect memory in these processes, since the most important actors are animals or people with memory.
To describe this type of behavior, we cannot use differential equations of integer orders and 
we need use mathematical tools that allow us to take into account the presence of memory in life, social and economic processes. 

The most important tool that allows us to describe processes with memory is the theory of integral-differential operators that form a calculus. Such operators are called the fractional derivatives and integrals, and their calculus is called fractional calculus \cite{FC1,FC2,FC3,FC4,FC5}. 
Fractional differential equations of non-integer orders with respect to time are a powerful tool for describing processes with memory in various sciences, including physics 
\cite{Handbook2019-4,Handbook2019-5}, economics 
\cite{BOOK-MDPI-2020,BOOK-DG-2021}, and other sciences. 

%%%%%%%%%%%%%%%%%%%%%%%%%%%%%%%%%%%%%%%%%%%%%%%%%%%%%%%%%%%%

The first attempt to account for memory in the predator-prey model was adopted in works \cite{Cavani-1,Cavani-2} in 1994. 
However, in these works, in fact, not memory was taken into account, but an exponentially distributed lag. 
Then, taking into account memory effects in predator-prey model, which is described by generalization of the Lotka-Volterra equations, was proposed in the works \cite{Das1,Das2} in 2009 and 2011 by using fractional differential equations of non-integer orders. 
In recent years, memory models have been actively investigated 
(for example, see \cite{Sahoo,Suryanto,Amirian,Tuladhar}). 
In all these papers, numerical simulations of the proposed predator-prey model are used. Exact solutions of generalized predator-prey model with memory are not proposed. 
 
%%%%%%%%%%%%%%%%%%%%%%%%%%%%%%%%%%%%%%%%%%%%%%%%%%%%%%%%%%%%
%%%%%%%%%%%%%%%%%%%%%%%%%%%%%%%%%%%%%%%%%%%%%%%%%%%%%%%%%%%%
%%%%%%%%%%%%%%%%%%%%%%%%%%%%%%%%%%%%%%%%%%%%%%%%%%%%%%%%%%%%

The approach to nonlinear dynamics, which is based on discrete maps, is important to study systems described by differential equations \cite{SUZ,Zaslavsky2,Chirikov,Schuster,CE}.
Discrete maps gives a much simpler formalism, which is useful in computer simulations.

In nonlinear dynamics and theory of deterministic chaos \cite{SUZ,Zaslavsky2,Chirikov,Schuster}, it is well-known that discrete maps can be derived from differential equations of integer orders with periodic kicks (for example, see Sections 5.2 and 5.3 in \cite[pp.60-68]{Zaslavsky2}, and Section 1.2 in \cite[pp.16-17]{Schuster}). In the proposed paper, we generalize this approach to derivation of the discrete maps for on the case of fading memory in a dynamical system.

The discrete maps with memory are considered, for example, in papers \cite{Ful,Fick1,Giona,Fick2,Gallas,StanMemory}.
Unfortunately, all these discrete maps with memory were not derived from any fractional differential equations. 

For the first time, discrete maps with memory were obtained from fractional differential equations in works 
\cite{Tarasov-Zaslavsky,Tarasov-Map1,Tarasov-Map2} 
(see also \cite{Springer2010} and \cite{Tarasov-Map3,TT-Logistic,Entropy2021}).
The proposed discrete maps with memory are exact discrete solutions of fractional differential equations with kicks.
No approximations were used when obtaining these maps with memory (for details see \cite[pp.409-453]{Springer2010}).

Then, the approach, which was suggested in \cite{Tarasov-Zaslavsky,Tarasov-Map1,Tarasov-Map2,Springer2010}, has been applied in works \cite{Tarasov-Map3,TT-Edelman1,TT-Edelman2,TT-Logistic,BOOK-DG-2021}. 
Computer simulations of some discrete maps with memory were suggested in papers \cite{Tarasov-Map3,TT-Edelman1,TT-Edelman2}, 
\cite{Edelman1,Edelman2,Edelman3,Edelman4,Edelman5}
\cite{Edelman6,Edelman7,Edelman8,Edelman-Handbook-2,Edelman-Handbook-4,Edelman-2021}, where new types of chaotic behavior and new kinds of attractors have been found. 

To obtain new type of discrete maps with memory from these equations, we use the equivalence of the fractional differential equation with Caputo fractional derivative and the Volterra integral equation of the second king (for details see \cite{Tarasov-Map1,Tarasov-Map2} and \cite{Springer2010}).
At the beginning, we obtain discrete memoryless maps from the first-order differential equation of Lotka-Volterra with periodic kicks in terms of interactions.
Then, we obtain the exact solutions for the fractional Lotka-Volterra and the fractional Kolmogorov equations with kicks, which takes into account power-law memory.
Using these exact solutions, we obtain discrete maps with memory for the proposed predator-prey models. 

%%%%%%%%%%%%%%%%%%%%%%%%%%%%%%%%%%%%%%%%%%%%%%%%%%%%%%%%%%%%

\section{Lotka-Volterra model with periodic kicks}

The Lotka-Volterra model is described by two first-order nonlinear differential equations. 
These equations are used in biology and economics 
to describe the dynamics of two species, which are called
a prey and a predator that interact with each other. 

The Lotka-Volterra equations have the form
\begin{equation} \label{LVE-X-1}
\frac{dX(t)}{dt} = a \, X(t) - b \, X(t) \, Y(t) ,
\end{equation}
\begin{equation} \label{LVE-Y-1}
\frac{dY(t)}{dt} = - c \, Y(t) + d \, X(t) \, Y(t) ,
\end{equation}
where $X(t)$ is the number of preys, and $Y(t)$ is the number of some predators.

The parameters $a$, $b$ $c$, $d$ are positive real numbers.
The prey are assumed to have an unlimited food supply and to reproduce exponentially, unless subject to predation. 
The parameter $a$ is the rate of this exponential growth.
The parameter $c$ represents the loss rate of the predators due to either natural death or emigration, 
it leads to an exponential decay in the absence of prey.

The parameters $b$ and $d$ are positive real parameters that characterize the interaction of the two species.
Let us assume that these interactions of the two species (prey and predator) occur for very short time interval, which can be considered as negligible time interval, almost instantaneous. 
We will assume that this interaction is realized in the form of sharp splashes ("kicks") that takes an infinitesimally short time. 
This property will be represented by series of the Dirac delta-functions.
We will assume that these sharp splashes ("kicks") occur after a certain time interval $T>0$. In the general case, this time interval can be considered as a random variable, and then the discrete maps can be averaged over some distribution. 
In our model, the parameter $T$ can be interpreted as the mean time between interactions. 
The parameters $b$ and $d$ characterize the intensity of this interaction, or rather the amplitudes of these sharp splashes ("kicks"). 

In this case, the Lotka-Volterra equations with periodic kicks can be represented in the form
\begin{equation} \label{LVE-X-2}
\frac{dX(t)}{dt} = a \, X(t) - b \, X(t) \, Y(t) \, \sum^{\infty}_{k=1} \delta \left(\frac{t}{T} -k \right) ,
\end{equation}
\begin{equation} \label{LVE-Y-2}
\frac{dY(t)}{dt} = - c \, Y(t) + d \, X(t) \, Y(t) \, \sum^{\infty}_{k=1} \delta \left(\frac{t}{T} -k \right) .
\end{equation}
In these equations, perturbations by interactions are taken into account by the periodic sequences of Dirac delta-function type kicks following with period $T$, where $b$ and $d$ are amplitudes of these kicks for prey and predator, respectively.

Equations \eqref{LVE-X-2} and \eqref{LVE-Y-2} can be written in the form
\begin{equation} \label{LVE-X-3}
\frac{d \ln X(t)}{dt} = a - b \, Y(t) \, \sum^{\infty}_{k=1} \delta \left(\frac{t}{T} -k \right) ,
\end{equation}
\begin{equation} \label{LVE-Y-3}
\frac{d \ln Y(t)}{dt} = - c + d \, X(t) \, \sum^{\infty}_{k=1} \delta \left(\frac{t}{T} -k \right) .
\end{equation}

Integration of \eqref{LVE-X-3} and \eqref{LVE-Y-3} from $0$ to $t$ gives
\begin{equation} \label{LVE-X-4}
\ln X(t) - \ln X(0)= a \, t - b \, \int^t_0 Y(\tau) \, \sum^{\infty}_{k=1} \delta \left(\frac{\tau}{T} - k \right) \, d\tau ,
\end{equation}
\begin{equation} \label{LVE-Y-4}
\ln Y(t) -\ln Y(0) = - c \, t + d \, \int^t_0 X(t) \, 
\sum^{\infty}_{k=1} 
\delta \left(\frac{\tau}{T} -k \right) \, d\tau .
\end{equation}
Let us use the equation
\begin{equation} \label{EQ-Delta}
\int^t_0 f(\tau) \, \delta \left(\frac{\tau}{T} -k \right) \, d\tau = T \, f(kT) ,
\end{equation}
which holds for $0<kT<t$, if $f(\tau)$ is continuous at $\tau=kT$. Then equations \eqref{LVE-X-4} and \eqref{LVE-Y-4} give
\begin{equation} \label{LVE-X-5}
\ln X(t) - \ln X(0)= a \, t - b \, T \, \sum^{\infty}_{k=1} Y(kT) \, \theta \left(t- k \, T \right) ,
\end{equation}
\begin{equation} \label{LVE-Y-5}
\ln Y(t) -\ln Y(0) = - c \, t + d \, T \, \sum^{\infty}_{k=1} X(kT) \, \theta \left(t- k \, T \right) ,
\end{equation}
where $\theta (t-kT)$ is the Heaviside step function, which is equal to zero when $k>t/T$ (i.e., $t<kT$).

We can see that solutions \eqref{LVE-X-5}, \eqref{LVE-Y-5} are discontinuous. The product of the delta-functions and the functions $X(t)$, $Y(t)$ in equations \eqref{LVE-X-2} and \eqref{LVE-Y-2}
is meaningful, if $X(t)$ and $Y(t)$ are continuous functions at the points $t=kT$. 
This is a contradiction, which can be resolved using $X(t-\varepsilon)$ and $Y(t-\varepsilon)$ with $0<\varepsilon <T$ ($\varepsilon \to 0+$) instead of $X(t)$ and $Y(t)$ to make a sense of these equations, when $X(kT-0) \neq X(kT+0)$ and $Y(kT-0) \neq Y(kT+0)$, \cite{Edelman3,Edelman6,Edelman7,Edelman8}. 

Expressions \eqref{LVE-X-5} and \eqref{LVE-Y-5} are exact solutions of the Lotka-Volterra equations with periodic kicks.

Let us derive the discrete-time maps by using solutions \eqref{LVE-X-5} and \eqref{LVE-Y-5}. 

For the left side of the $(n+1)$-th kicks, where $nT <t<(n+1)T$, 
equation \eqref{LVE-X-5} and \eqref{LVE-Y-5} can be written in the form 
\begin{equation} \label{LVE-X-5b}
\ln X(t) - \ln X(0)= 
a \, t - b \, T \, \sum^{n}_{k=1} Y_k ,
\end{equation}
\begin{equation} \label{LVE-Y-5b}
\ln Y(t) -\ln Y(0) = 
- c \, t + d \, T \, \sum^{n}_{k=1} X_k ,
\end{equation}
where $X_k$ and $Y_k$ are defined by the expressions
\begin{equation} \label{XYn}
X_{k}=\lim_{\varepsilon \rightarrow 0+} X(kT-\varepsilon) , 
\quad
Y_{k}=\lim_{\varepsilon \rightarrow 0+} Y(kT-\varepsilon) ,
\end{equation}
for $k \in \mathbb{N}$.

For $t =(n+1)T-\varepsilon$ at $\varepsilon \to 0+$ in equations \eqref{LVE-X-5b} and \eqref{LVE-Y-5b}, we get 
\begin{equation} \label{LVE-X-6}
\ln X_{n+1} - \ln X_0 = 
a \, (n+1) \, T - b \, T \, \sum^{n}_{k=1} Y_k , 
\end{equation}
\begin{equation} \label{LVE-Y-6}
\ln Y_{n+1} -\ln Y_0 = 
- c \, (n+1) \, T + d \, T \, \sum^{n}_{k=1} X_k .
\end{equation}

For $t =nT-\varepsilon$ at $\varepsilon \to 0+$ in equations \eqref{LVE-X-5b} and \eqref{LVE-Y-5b}, we get
\begin{equation} \label{LVE-X-7}
\ln X_n - \ln X_0 = 
a \, n \, T - b \, T \, \sum^{n-1}_{k=1} Y_k , 
\end{equation}
\begin{equation} \label{LVE-Y-7}
\ln Y_n -\ln Y_0 = 
- c \, n \, T + d \, T \, \sum^{n-1}_{k=1} X_k .
\end{equation}

Subtracting equation \eqref{LVE-X-7} and \eqref{LVE-Y-7} 
of the $n$-th step from equations \eqref{LVE-X-6} 
\eqref{LVE-Y-6} of the $(n+1)$-th step, 
we obtain the equations of the discrete-time maps
\begin{equation} \label{LVE-X-8}
\ln X_{n+1} - \ln X_{n} = a \, T - b \, T \, Y_n , 
\end{equation}
\begin{equation} \label{LVE-Y-8}
\ln Y_{n+1} -\ln Y_{n} = - c \, T + d \, T \, X_n .
\end{equation}

These maps can be rewritten in the form
\begin{equation} \label{LVE-X-9}
X_{n+1} = X_{n} \exp \{ - (b \, Y_n -a) \, T \} , 
\end{equation}
\begin{equation} \label{LVE-Y-9}
Y_{n+1} = Y_{n} \exp \{ - ( d \, X_n -c) \, T \}.
\end{equation}

Discrete maps \eqref{LVE-X-9} and \eqref{LVE-Y-9} describe behaviors of predator and prey without memory
in the framework of proposed generalization of the Lotka-Volterra model that is described by equations
\eqref{LVE-X-2} and \eqref{LVE-Y-2}. 

%%%%%%%%%%%%%%%%%%%%%%%%%%%%%%%%%%%%%%%%%%%%%%%%%%%%%%
%%%%%%%%%%%%%%%%%%%%%%%%%%%%%%%%%%%%%%%%%%%%%%%%%%%%%%
%%%%%%%%%%%%%%%%%%%%%%%%%%%%%%%%%%%%%%%%%%%%%%%%%%%%%%

\section{Lotka-Volterra model with kicks and memory}

The description of life, social and economic processes should take into account a memory i.e. that the behavior of actors may depend on the history of previous changes in processes.
It is strange to neglect memory in these processes, since the most important actors are animals and people, which have a memory.
To describe this type of behavior, we cannot use differential equations of integer orders and 
we need use mathematical tools that allow us to take into account the presence of memory in life, social and economic processes. 

To take into account a fading memory, we can consider integro-differential equations instead of differential equations of the integer orders. 
If we take into account the memory with power-law fading and periodic kicks, then we get the generalization of Lotka-Volterra model represented by the integro-differential equations
\begin{equation} \label{LVE-X-3M}
\int^{t}_0 d\tau \, M_X(t,\tau) \, \left( \frac{d \ln X(\tau)}{d\tau} \right) =
a - b \, Y(t) \, \sum^{\infty}_{k=1} \delta \left(\frac{t}{T} - k \right) ,
\end{equation}
\begin{equation} \label{LVE-Y-3M}
\int^{t}_0 d\tau \, M_Y(t,\tau) \, \left( \frac{d \ln Y(\tau)}{d\tau} \right) = - c + d \, X(t) \, \sum^{\infty}_{k=1} \delta \left(\frac{t}{T} - k \right) ,
\end{equation}
where $M_X(t,\tau)$ and $M_Y(t,\tau)$ are memory functions.
The properties of this functions are described, for example, in \cite{Concept} and \cite[pp.3-52]{BOOK-DG-2021}.
For $M_X(t,\tau)=M_Y(t,\tau) = \delta (t-\tau)$, 
equations \eqref{LVE-X-3M} and \eqref{LVE-Y-3M} give equations 
\eqref{LVE-X-2} and \eqref{LVE-Y-2} that described dynamics without memory.

We can consider the power-law form of memory fading and powe-law memory functions due to the following reasons. 

(I) Power laws and power-law functions play an important role in different life and social sciences \cite{Pinto}:
\begin{enumerate}
\item Biology \cite{Pinto,Schuster2011,Torres};
\item Ecology \cite{Pinto,Marquet};
\item Population dynamics \cite{Pinto,Luscombe};
\item Victimology \cite{Pinto,Pinto2};
\item Economics \cite{Pinto,Gabaix2,Gabaix1};
\item Finance \cite{Gabaix1}.
\end{enumerate}

(II) The power-law memory function
can be considered as an approximation of the generalized memory functions.
In paper \cite{General}, we use the generalized Taylor series in the Trujillo-Rivero-Bonilla form for wide class of the memory functions. We proved that the equations with memory functions can be represented through the integro-differential operators with power-law kernels.

As a result, we will assume the following power-law form of memory function 
\begin{equation} \label{EQ-MX} 
M_X(t,\tau) =
\frac{1}{\Gamma (1-\alpha)} (t-\tau)^{-\alpha}, 
\end{equation} 
\begin{equation} \label{EQ-MY} 
M_Y(t,\tau) =
\frac{1}{\Gamma (1-\beta)} (t-\tau)^{-\beta}, 
\end{equation} 
where $\Gamma (z)$ is the gamma function, $\alpha>0$ and $\beta>0$ are the memory fading parameters for behavioral processes of prey and predator, respectively.

%%%%%%%%%%%%%%%%%%%%%%%%%%%%%%%%

If we take into account the memory with power-law fading by the
memory functions \eqref{EQ-MX} and \eqref{EQ-MY}, then 
the generalized Lotka-Volterra equations with kicks and memory 
\eqref{LVE-X-3M} and \eqref{LVE-Y-3M} take the form
\begin{equation} \label{LVE-X-3F}
\left( D^{\alpha}_{C,0+} \ln X \right) (t) =a - b \, Y(t) \, \sum^{\infty}_{k=1} \delta \left(\frac{t}{T} - k \right) ,
\end{equation}
\begin{equation} \label{LVE-Y-3F}
\left( D^{\beta}_{C,0+} \ln Y \right)(t) = - c + d \, X(t) \, \sum^{\infty}_{k=1} \delta \left(\frac{t}{T} - k \right) ,
\end{equation}
For $\alpha=\beta=1$, equations \eqref{LVE-X-3F} and \eqref{LVE-Y-3F} gives \eqref{LVE-X-2} and \eqref{LVE-Y-2}.

The Caputo fractional derivative is defined by the equation
\begin{equation} \label{EQ-Caputo} 
\left(D^{\alpha}_{C;a+} f\right)(t) = 
\frac{1}{\Gamma (N-\alpha)}
\int^{t}_a (t-\tau)^{N-\alpha -1} f^{(N)}(\tau) d\tau , 
\end{equation} 
where $t \in [a,b]$, and $f^{(N)} (\tau)$ is the derivative of the integer order $N$ with $N-1<\alpha\le N$. In equation \eqref{EQ-Caputo}, it is assumed that $f(\tau) \in AC^{N}[a,b]$, i.e., the function $f(\tau)$ has integer-order derivatives up to ($N-1$)-th order, which are continuous functions on the interval $[a,b]$, and the derivative $f^{(N)}(\tau)$ is Lebesgue summable on the interval $[a,b]$.

To obtain new type of discrete maps with memory from these equations, we the equivalence of the fractional differential equation with Caputo fractional derivative and the Volterra integral equation of the second king (for details see \cite{Springer2010}).

To get the solution, we will use the equivalence of the fractional differential equation with Caputo fractional derivative and the Volterra integral equation of the second king (for details see \cite{Tarasov-Map1,Tarasov-Map2} and \cite{Springer2010}). This equivalence is based on the equality
(see Lemma 2.22 in \cite[p.~96]{FC4}) in the from
\begin{equation} \label{EQ-FTFC} 
\left(I^{\alpha}_{RL;a+} D^{\alpha}_{C;a+}f \right)(t)=
f(t)-\sum^{N-1}_{k=0} \frac{f^{(k)}(a+)}{k!} \, (t-a)^k , 
\end{equation} 
which holds for $f(t)\in AC^{N}[a,b]$ or $f(t)\in C^{N}[a,b]$, 
where $N<\alpha \le N$. 
For the case $\alpha \in (0,1)$, $(N=1)$, equation \eqref{EQ-FTFC} takes the form
\begin{equation} \label{EQ5-4_30} 
\left(I^{\alpha}_{RL;a+} \, D^{\alpha}_{C;a+}f \right)(t) = 
f(t)-f(a) 
\end{equation} 
for $f(t)\in AC[a,b]$ or $f(t)\in C[a,b]$.
Here $I^{\alpha}_{RL;a+}$ is the Riemann-Liouville fractional integral that is defined by the equation
\begin{equation} \label{EQ1-27} 
\left(I^{\alpha}_{RL;a+} f \right)(t) = \frac{1}{\Gamma (\alpha)}\int^t_a (t-\tau)^{\alpha-1} f (\tau) d\tau , 
\end{equation} 
where $\Gamma (\alpha)$ is the gamma function, and the function $f(t)$ satisfies the condition $f(t) \in L_1(a,b)$.
For $\alpha =1$, operator \eqref{EQ1-27} gives the standard integration of the first order 
\begin{equation}
\left(I^1_{RL;a+} f\right)(t) = \int^t_a f (\tau) \, d\tau .
\end{equation}

Let us consider the case $\alpha,\beta \in (0,1)$. 
Then, application of the Riemann-Liouville fractional integrals of orders alpha and beta, respectively, gives the equations 
\begin{equation} \label{LVE-X-3F2}
\ln X (t) - \ln X(0) = a \left( I^{\alpha}_{RL,0+} 1 \right)(t) - 
b \, \left( I^{\alpha}_{RL,0+} \, Y(\tau) \, \sum^{\infty}_{k=1} \delta \left(\frac{\tau}{T} - k \right) \right) (t) ,
\end{equation}
\begin{equation} \label{LVE-Y-3F2}
\ln Y(t) - \ln Y(0) = - c \left( I^{\beta}_{RL,0+} 1 \right)(t) + 
d \, \left( I^{\beta}_{RL,0+} X(\tau) \, \sum^{\infty}_{k=1} \delta \left(\frac{\tau}{T} - k \right) \right) (t),
\end{equation}

Fractional differential equations \eqref{LVE-X-3F2} and \eqref{LVE-Y-3F2} contain the Dirac delta-functions. These functions are the generalized functions \cite{GenFunc1,GenFunc2} that are treated as functionals on a space of test functions. 
Therefore equations \eqref{LVE-X-3F2} and \eqref{LVE-Y-3F2} should be considered in a generalized sense, i.e. on the space of test functions, which are continuous. 
The product of the delta-functions and the functions $X(t)$, $Y(t)$ is meaningful, if $X(t)$ and $Y(t)$ are continuous functions at the points $t=kT$. 
Therefore we use $X(t-\varepsilon)$ and $Y(t-\varepsilon)$ with 
$0<\varepsilon <T$ ($\varepsilon \to 0+$) instead of $X(t)$ and $Y(t)$ to make a sense of these equations for $\alpha, \beta \in (0,1)$, when $X(kT-0) \neq X(kT+0)$ and $Y(kT-0) \neq Y(kT+0)$, \cite{Edelman3,Edelman6,Edelman7,Edelman8}.

For $t \in (nT,(n+1)T)$ in equations \eqref{LVE-X-5} and \eqref{LVE-Y-5}, we get 
\begin{equation} \label{LVE-X-3F3}
\ln X (t) - \ln X(0) = a \, \left( I^{\alpha}_{RL,0+} 1 \right)(t) - 
\frac{b \, T}{\Gamma (\alpha)} \, \sum^{n}_{k=1} Y(kT) \, (t-kT)^{\alpha-1} \, \theta \left(t-k T \right) ,
\end{equation}
\begin{equation} \label{LVE-Y-3F3}
\ln Y(t) - \ln Y(0) = - c \, \left( I^{\beta}_{RL,0+} 1 \right)(t) + 
\frac{d \, T}{\Gamma (\beta)} \, \sum^{n}_{k=1} X(kT) \, (t-kT)^{\alpha-1} \, \theta \left(t-k T \right) .
\end{equation}

Using equation 2.1.16 of \cite{FC4} (see Property 2.1 in \cite[p.71]{FC4}) for the Riemann-Liouville fractional integral
\begin{equation}
\left( I^{\alpha}_{RL,0+} t^{\delta} \right)(t) =
\frac{\Gamma (\delta+1)}{\Gamma(\alpha+\delta+1)} 
\, t^{\alpha+\delta} ,
\end{equation}
where $\delta>-1$ and $\alpha>0$,
we get equations \eqref{LVE-X-3F3} and \eqref{LVE-Y-3F3} in the form
\begin{equation} \label{LVE-X-3F4}
 \ln X(t) - \ln X(0) = \frac{a}{\Gamma(\alpha+1)} t^{\alpha} - 
\frac{b \, T}{\Gamma (\alpha)} \, \sum^{n}_{k=1} Y(kT) \, (t-kT)^{\alpha-1} \, \theta \left(t-k T \right) ,
\end{equation}
\begin{equation} \label{LVE-Y-3F4}
\ln Y(t) - \ln Y(0) = - \frac{c}{\Gamma(\beta+1)} t^{\beta} - 
\frac{d \, T}{\Gamma (\beta)} \, \sum^{n}_{k=1} X(kT) \, (t-kT)^{\beta-1} \, \theta \left(t-k T \right) .
\end{equation}

Expressions \eqref{LVE-X-3F3} and \eqref{LVE-Y-3F3} are the exact solutions of the generalized Lotka-Volterra model with power-law memory and periodic kicks \eqref{LVE-X-3F}, \eqref{LVE-Y-3F}.

Let us derive the discrete-time maps for these solutions. 

For the left side of the $k$-th kicks ($t=kT$), 
we can define $X_k$ and $Y_k$ by equation \eqref{XYn}.
For $t =(n+1)T-\varepsilon$ at $\varepsilon \to 0+$ in equations \eqref{LVE-X-3F4} and \eqref{LVE-Y-3F4}, we get
\begin{equation} \label{LVE-X-3F5}
 \ln X_{n+1} - \ln X_0 = \frac{a \, T^{\alpha}}{\Gamma(\alpha+1)} (n+1)^{\alpha} - 
\frac{b \, T^{\alpha}}{\Gamma (\alpha)} \, \sum^{n}_{k=1} Y_k \, (n+1-k)^{\alpha-1} ,
\end{equation}
\begin{equation} \label{LVE-Y-3F5}
\ln Y_{n+1} - \ln Y_0 = - \frac{c \, T^{\beta}}{\Gamma(\beta+1)} (n+1)^{\beta} - 
\frac{d \, T^{\beta}}{\Gamma (\beta)} \, \sum^{n}_{k=1} X_k \, (n+1-k)^{\beta-1} .
\end{equation}
For $t =nT-\varepsilon$ at $\varepsilon \to 0+$ in equations \eqref{LVE-X-3F4} and \eqref{LVE-Y-3F4}, we get
\begin{equation} \label{LVE-X-3F6}
 \ln X_{n} - \ln X_0 = \frac{a \, T^{\alpha}}{\Gamma(\alpha+1)} (n)^{\alpha} - 
\frac{b \, T^{\alpha}}{\Gamma (\alpha)} \, \sum^{n-1}_{k=1} Y_k \, (n-k)^{\alpha-1} ,
\end{equation}
\begin{equation} \label{LVE-Y-3F6}
\ln Y_{m} - \ln Y_0 = - \frac{c \, T^{\beta}}{\Gamma(\beta+1)} (n)^{\beta} - 
\frac{d \, T^{\beta}}{\Gamma (\beta)} \, \sum^{n-1}_{k=1} X_k \, (n-k)^{\beta-1} .
\end{equation}

Subtracting equation \eqref{LVE-X-3F6} and \eqref{LVE-Y-3F6}
of the $n$-th step from equation \eqref{LVE-X-3F5} 
\eqref{LVE-Y-3F5} for the $(n+1)$-th step, 
we obtain the equations of the discrete-time maps with memory
\begin{equation} \label{LVE-X-3F7}
 \ln X_{n+1} - \ln X_{n} = \frac{a \, T^{\alpha}}{\Gamma(\alpha+1)} {\cal V}_{\alpha+1} (n)
%%%\left( (n+1)^{\alpha} - n^{\alpha} \right) 
- \frac{b \, T^{\alpha}}{\Gamma (\alpha)} \, Y_n +
\frac{b \, T^{\alpha}}{\Gamma (\alpha)} \, \sum^{N-1}_{k=1} Y_k \, {\cal V}_{\alpha} (n-k) ,
%%%\left( (n+1-k)^{\alpha-1} - (n-k)^{\alpha-1} \right) ,
\end{equation}
\begin{equation} \label{LVE-Y-3F7}
\ln Y_{n+1} - \ln Y_{n} = - \frac{c \, T^{\beta}}{\Gamma(\beta+1)} {\cal V}_{\beta+1} (n)
%%% \left( (n+1)^{\beta} - m^{\beta} \right) 
+ \frac{d \, T^{\beta}}{\Gamma (\beta)} \, X_{n} +
\frac{d \, T^{\beta}}{\Gamma (\beta)} \, \sum^{N-1}_{k=1} X_k \, {\cal V}_{\beta} (n-k)
%%% \left( (n+1-k)^{\beta-1} - (n-k)^{\beta-1} \right) . 
\end{equation}
where we use the function
\be
{\cal V}_{\alpha} (z) = (z+1)^{\alpha-1} - z^{\alpha-1} 
\ee
for $z > 0$.

Discrete maps \eqref{LVE-X-3F7} and \eqref{LVE-Y-3F7} describe behaviors of predator and prey with power-law fading memory
in the framework of generalized Lotka-Volterra model with periodic kicks and power-law memory.

%%%%%%%%%%%%%%%%%%%%%%%%%%%%%%%%%%%%%%%%%%%%%%%%%%%%%%%%%%
%%%%%%%%%%%%%%%%%%%%%%%%%%%%%%%%%%%%%%%%%%%%%%%%%%%%%%%%%%
%%%%%%%%%%%%%%%%%%%%%%%%%%%%%%%%%%%%%%%%%%%%%%%%%%%%%%%%%%

\section{Kolmogorov model with periodic kicks and memory}

Let us consider economic model of two interacting sectors (or two economic processes) \cite[pp.377-381]{BOOK-DG-2021}, \cite{CNSNS2019} that is described by two first-order differential equations
\begin{equation} \label{EQ-PP-1a} 
\frac{dX(t)}{dt} = X(t) \, F_1(X(t),Y(t)) ,
\end{equation} 
\begin{equation} \label{EQ-PP-1b} 
\frac{dY(t)}{dt} = Y(t) \, F_2(X(t),Y(t)) .
\end{equation}
The functions $F_1$ and $F_2$ are interpreted as the respective growth rates of these two sectors.
Equations \eqref{EQ-PP-1a} and \eqref{EQ-PP-1b} define the Kolmogorov predator-prey model.

We should note that these equations can be written in the form
\begin{equation} \label{EQ-PP-2a} 
\frac{d \ln X(t)}{dt} = F_1(X(t),Y(t)) ,
\end{equation} 
\begin{equation} \label{EQ-PP-2b} 
\frac{d \ln Y(t)}{dt} = F_2(X(t),Y(t)) .
\end{equation}

Let us consider a generalization of the Kolmogorov predator-prey model by taking into account periodic kicks.
Equations with periodic kicks have the form
\begin{equation} \label{EQ-PP-3a} 
\frac{d \ln X(t)}{dt} = F_1(X(t),Y(t)) \,
\sum^{\infty}_{k=1} \delta \left(\frac{t}{T} - k \right) 
\end{equation} 
\begin{equation} \label{EQ-PP-3b} 
\frac{d \ln Y(t)}{dt} = F_2(X(t),Y(t)) \,
\sum^{\infty}_{k=1} \delta \left(\frac{t}{T} - k \right) .
\end{equation}

The predator-prey model with periodic kicks and power-law memory is described by the fractional differential equations
\begin{equation} \label{EQ-PP-F1a} 
\left(D^{\alpha_1}_{C;0+} \ln X\right)(t) = F_1(X(t),Y(t)) \,
\sum^{\infty}_{k=1} \delta \left(\frac{t}{T} - k \right) 
\end{equation} 
\begin{equation} \label{EQ-PP-F1b} 
\left(D^{\alpha_2}_{C;0+} \ln Y\right)(t) = F_2(X(t),Y(t)) \,
\sum^{\infty}_{k=1} \delta \left(\frac{t}{T} - k \right) .
\end{equation}
For $\alpha_1=\alpha_2=1$, equations \eqref{EQ-PP-F1a} and \eqref{EQ-PP-F1b} take the form of equations 
\eqref{EQ-PP-1a} and \eqref{EQ-PP-1b}.

In equations \eqref{EQ-PP-F1a} and \eqref{EQ-PP-F1b},
the products of the delta-functions and the functions $F_1(X(t),Y(t))$, $F_2(X(t),Y(t))$, are meaningful, if $F_1(X(t),Y(t))$,
$F_2(X(t),Y(t))$ are continuous functions at the points $t=kT$. 
Therefore, we use the functions 
\begin{equation}
F_{s,\varepsilon}(X(t),Y(t)) = 
F_{s}(X(t-\varepsilon),Y(t-\varepsilon)) , \quad (s=1,2)
\end{equation}
with $0<\varepsilon <T$ ($\varepsilon \to 0+$) instead of $F_s(X(t),Y(t))$, $s=1,2$ to make a sense of these equations, when $X(kT-0) \neq X(kT+0)$ and $Y(kT-0) \neq Y(kT+0)$, \cite{Edelman3,Edelman6,Edelman7,Edelman8}. 

%%%%%%%%%%%%%%%%%%%%%%%%%%%%%%%%%%%%%%%%%%%%%%%%%%%%%%%%%%
%%%%%%%%%%%%%%%%%%%%%%%%%%%%%%%%%%%%%%%%%%%%%%%%%%%%%%%%%%
%%%%%%%%%%%%%%%%%%%%%%%%%%%%%%%%%%%%%%%%%%%%%%%%%%%%%%%%%%

By repeating the transformations of the previous section, we can get the exact solution of these fractional differential equations and discrete mappings with memory. 

Let us consider the case $\alpha_1,\alpha_2 \in (0,1)$. 
Then, application of the Riemann-Liouville fractional integrals of orders $\alpha_1$ and $\alpha_2$, respectively, gives the equations 
\begin{equation} \label{KM-X-1}
\ln X (t) - \ln X(0) = 
\left( I^{\alpha_1}_{RL,0+} \, F_2(X(\tau),Y(\tau)) 
 \, \sum^{\infty}_{k=1} \delta \left(\frac{\tau}{T} - k \right) \right) (t) ,
\end{equation}
\begin{equation} \label{KM-Y-1}
\ln Y(t) - \ln Y(0) = 
\left( I^{\alpha_2}_{RL,0+} \, F_1(X(\tau),Y(\tau)) \, 
\sum^{\infty}_{k=1} \delta \left(\frac{\tau}{T} - k \right) \right) (t) .
\end{equation}

Fractional differential equations \eqref{KM-X-1} and \eqref{KM-Y-1} contain the Dirac delta-functions. 
The product of the delta-functions and $F_k(X(\tau),Y(\tau))$ is meaningful, if the functions $F_k(X(\tau),Y(\tau))$ are continuous at the points $t=kT$. 
Therefore we use $F_k(X(t-\varepsilon), Y(t-\varepsilon))$ with $0<\varepsilon <T$ ($\varepsilon \to 0+$) instead of $F_k(X(\tau),Y(\tau))$ to make a sense of these equations for $\alpha_k \in (0,1)$, when $X(kT-0) \neq X(kT+0)$ and $Y(kT-0) \neq Y(kT+0)$, \cite{Edelman3,Edelman6,Edelman7,Edelman8}. 

For $t \in (nT,(n+1)T)$ at $\varepsilon \to 0+$ in equations 
\eqref{KM-X-1} and \eqref{KM-Y-1}, we get
\begin{equation} \label{KM-X-2}
\ln X (t) - \ln X(0) = 
\frac{T}{\Gamma (\alpha_1)} \, \sum^{n}_{k=1} F_1(X(kT),Y(kT)) \, (t-kT)^{\alpha_1-1} \, \theta \left(t-k T \right) ,
\end{equation}
\begin{equation} \label{KM-Y-2}
\ln Y (t) - \ln Y(0) = 
\frac{T}{\Gamma (\alpha_2)} \, \sum^{n}_{k=1} F_2(X(kT),Y(kT)) \, (t-kT)^{\alpha_2-1} \, \theta \left(t-k T \right) .
\end{equation}

Expressions \eqref{KM-X-2} and \eqref{KM-Y-2} describe solutions 
of equations of the generalized Kolmogorov model with power-law memory and periodic kicks.
Let us derive the discrete-time maps for these solutions
by using $X_n$ and $Y_n$ that are defined by equation \eqref{XYn}.

For $t =(n+1)T-\varepsilon$ at $\varepsilon \to 0+$ in equations 
\eqref{KM-X-2} and \eqref{KM-Y-2}, we get 
\begin{equation} \label{KM-X-3}
 \ln X_{n+1} - \ln X_0 = 
\frac{T^{\alpha_1}}{\Gamma (\alpha_1)} \, 
\sum^{n}_{k=1} F_1(X_k,Y_k) \, (n+1-k)^{\alpha_1-1} ,
\end{equation}
\begin{equation} \label{KM-Y-3}
\ln Y_{n+1} - \ln Y_0 = 
\frac{T^{\alpha_2}}{\Gamma (\alpha_2)} \, 
\sum^{n}_{k=1} F_2(X_k,Y_k) \, (n+1-k)^{\alpha_2-1} .
\end{equation}

For $t =nT-\varepsilon$ at $\varepsilon \to 0+$, we have
\begin{equation} \label{KM-X-4}
 \ln X_{n} - \ln X_0 = 
\frac{T^{\alpha}}{\Gamma (\alpha_1)} \, 
\sum^{n-1}_{k=1} F_1(X_k,Y_k) \, (n-k)^{\alpha_1-1} ,
\end{equation}
\begin{equation} \label{KM-Y-4}
\ln Y_{n} - \ln Y_0 = 
\frac{T^{\alpha_2}}{\Gamma (\alpha_2)} \, 
\sum^{n-1}_{k=1} F_2(X_k,Y_k) \, (n-k)^{\alpha_2-1} .
\end{equation}

Subtracting equations \eqref{KM-X-4} and \eqref{KM-Y-4} of the $n$-th step from equations \eqref{KM-X-3} and \eqref{KM-Y-3} of the $(n+1)$-th step, we obtain the equations of the discrete maps with memory
\begin{equation} \label{KM-X-5}
 \ln X_{n+1} - \ln X_n = 
\frac{T^{\alpha_1}}{\Gamma (\alpha_1)} \, F_1(X_n,Y_n) +
\frac{T^{\alpha_1}}{\Gamma (\alpha_1)} \, 
\sum^{n-1}_{k=1} F_1(X_k,Y_k) \, {\cal V}_{\alpha_1} (n-k) ,
\end{equation}
\begin{equation} \label{KM-Y-5}
\ln Y_{n+1} - \ln Y_n = 
\frac{T^{\alpha_2}}{\Gamma (\alpha_2)} \, F_2(X_n,Y_n) +
\frac{T^{\alpha_2}}{\Gamma (\alpha_2)} \, 
\sum^{n-1}_{k=1} F_2(X_k,Y_k) \, {\cal V}_{\alpha_2} (n-k) ,
\end{equation}
where we use the function
\be
{\cal V}_{\alpha} (z) = (z+1)^{\alpha-1} - z^{\alpha-1} 
\ee
for $z > 0$.

Discrete maps \eqref{KM-X-5} and \eqref{KM-Y-5} describe behaviors of predator and prey with power-law fading memory
in the framework of generalized Kolmogorov model with periodic kicks and power-law memory. 

%%%%%%%%%%%%%%%%%%%%%%%%%%%%%%%%%%%%%%%%%%%%%%%%%%%%%%%%%%

We can consider a special case of the proposed general model.
For example, we can use the expansion of the functions $F_k(X(t),Y(t))$ of two variables in the Taylor series. 
We can consider only the linear case
\begin{equation}
F_k(X,Y)=\lambda_k +a_kX+ b_kY ,
\end{equation}
where
\begin{equation}
a_k=\left( \frac{F_k(X,Y)}{\partial X} \right)_{X=Y=0} , \quad
b_k=\left( \frac{F_k(X,Y)}{\partial Y} \right)_{X=Y=0} .
\end{equation}
If we assume that $a_1=0$, $b_2=0$
and relabel the parameters $a_k$ and $b_k$ such that
\begin{equation} \label{param}
\lambda_1=a , \quad \lambda_2=-c , \quad
a_2=d , \quad b_1=-b ,
\end{equation}
then equations \eqref{EQ-PP-F1a} and \eqref{EQ-PP-F1b} give the equations of the new generalized Lotka-Volterra model with memory and kicks. 
Note that the first model, which is described by equations \eqref{LVE-X-3F} and \eqref{LVE-Y-3F}, the linear terms with parameters 
$\lambda_1=a$ and $\lambda_2=-c$ are outside the action of periodic kicks. 
Therefore, in case \eqref{param} with $a_1=b_2=0$, equations \eqref{KM-X-5} and \eqref{KM-Y-5} do not coincide with equations of discrete maps with memory \eqref{LVE-X-3F7} and \eqref{LVE-Y-3F7}.

%%%%%%%%%%%%%%%%%%%%%%%%%%%%%%%%%%%%%%%%%%%%%%%%%%%%%%%%%%

\section{Conclusion}

In this paper, new type of predator-prey models is proposed.
These models take into account power-law fading memory and periodic kicks. 
The equations of these models are generalizations of the Lotka-Volterra and Kolmogorov equations, in which we take into account power-law memory by using the fractional derivatives of non-integer orders and periodic kicks. 
This derivative allows us to take into account memory with power-law fading.
The periodic kicks take into account short duration of interaction between predators and prey.
In this paper, we obtain exact solutions that describe behaviors of predator and prey with power-law fading memory. 
These exact solutions are used to get discrete maps with memory that represent the proposed predator-prey models with memory. 
It should be emphasized that these discrete memory maps with memory are obtained from fractional differential equations without using any approximations. 

In this paper, we proposed models and discrete maps with memory for the case $\alpha,\beta \in (0,1)$ and $\alpha_1,\alpha_2 \in(0,1)$ to simplify consideration and equations. 
These models can be simply generalized for all positive values of these parameters by using the methods, which are described in Chapter 18 of \cite[pp.409-453]{Springer2010}. 
For this purpose it is necessary to use generalized moments for variables $X(t)$, $Y(t)$, and Theorems 18.17-18.19 of 
\cite[pp.442-445]{Springer2010}.

We assume that proposed predator-prey models 
and discrete maps with memory will find many applications in the nonlinear economic and biological dynamics with memory. 

It is safe to hope that the proposed maps with memory can simplify simulations of the behavior of prey and predators with power-law fading memory in computer simulations. However, this modeling remains an open question hopefully it will be solved in future research. 

%%%%%%%%%%%%%%%%%%%%%%%%%%%%%%%%%%%%%%%%%%%%%%%%%%%%%%%%%%

\section*{conflicts of Interest Statement}

This work does not have any conflicts of interest.

%%%%%%%%%%%%%%%%%%%%%%%%%%%%%%%%%%%%%%%%%%%%%%%%%%%%%%%%%%

%%%%%%%%%%%%%%%%%%%%%%%%%%%%%%%%%%%%%%%%%%%%%%%%%%%%%%%%%%

\section*{Funding Statemen}

There are no funders to report for this submission

%%%%%%%%%%%%%%%%%%%%%%%%%%%%%%%%%%%%%%%%%%%%%%%%%%%%%%%%%%
%%%%%%%%%%%%%%%%%%%%%%%%%%%%%%%%%%%%%%%%%%%%%%%%%%%%%%%%%%
%%%%%%%%%%%%%%%%%%%%%%%%%%%%%%%%%%%%%%%%%%%%%%%%%%%%%%%%%%

%%%%%%%%%%%%%%%%%%%%%%%%%%%%%%%%%%%%%%%%%%%%%%%%%%%%%%%%%%
%%%%%%%%%%%%%%%%%%%%%%%%%%%%%%%%%%%%%%%%%%%%%%%%%%%%%%%%%%
%%%%%%%%%%%%%%%%%%%%%%%%%%%%%%%%%%%%%%%%%%%%%%%%%%%%%%%%%%

\end{document}